# Quasilattices of the Spectre monotile

Henning U. Voss[1,2] and Douglas J. Ballon[2]

[1]*Cornell MRI Facility, Cornell University, Ithaca, NY, USA.* [2]*Department of Radiology, Weill Cornell Medicine, New York, NY, USA*

February 12, 2025

The Spectre is a family of recently discovered aperiodic monotiles that tile the plane only in non-periodic ways, and novel physical phenomena have been predicted for planar systems made of aperiodic monotiles. It is shown that point decorations of Tile(1,1), the base tile for all Spectres, supports the generation of a large variety of non-periodic quasilattices, in contrast to Bravais-lattices in which all point decorations would be periodic. A lattice generating function is introduced as a mapping from point decorations to quasilattice space, and investigated systematically. It is found that some lattices result from the properties of nearest-neighbor distances of point decorations, and that other lattices show near-periodicity in projections along one of the symmetry axes of the tiling. It is concluded that the lattice generating function can serve as a template for the design of physical potential landscapes that can be controlled by the point decoration as a parameter.

Recently, Smith et al. [1] discovered a family of aperiodic monotiles of the plane, Tile($a,b$), with geometric parameters *a* and *b*. In particular, the special case of Tile(1,1) stands out, as it is a weakly chiral monotile; in other words, Tile(1,1) can tile the plane only in a non-periodic fashion, and reflected (turned-over) tiles are neither needed nor allowed. A relative of Tile(1,1), the so-called *Hat*, is a non-chiral monotile and requires the inclusion of reflected tiles for the non-periodic tiling of the plane [2]. A generalization of Tile(1,1), the *Spectre*, consists of a family of Tile(1,1) monotiles with deformed edges. The Spectre is a strictly chiral aperiodic monotile, and reflected tiles are allowed but do not contribute to the tilings [1, 3].

Figure 1 shows one possibility of a *decoration*, or marker, that is repeated identically on all tiles. We use the term decoration here as it is often defined in condensed matter physics [4], without the implied notion that decorations impose certain rules on tile matching. Here the focus of attention is put on the *point set* defined by point decorations, which consist of one selected point on the tile. Physically, single-point decorations could be interpreted as the center of a potential associated with each of the tiles, and disc-shaped decorations could indicate the spatial extent of a potential.

Edges themselves are not considered of importance here, for various reasons. (i) One can always define the point set by selecting a decoration with a point on one of the edges. (ii) In physical systems such as crystals, it is the point set that defines the potential locations and not the edge of a unit cell. (iii) Specifically for the considered Tile(1,1), the edges can be deformed quite generally, leading to the Spectre tiling [1]. Thus, the Spectre tiling is considered here physically equivalent to the Tile(1,1) tiling. (iv) Decorations do not need to be restricted to the tile area. For example, a molecule located at the center of a tile could have a potential or binding site that extends beyond the tile edge. (v) A point decoration that is located outside of Tile(1,1) might still be inside an equivalent Spectre tile, depending on its edge definition.

To systematically investigate point decorations, we define a generating function $P: \Omega \to \Sigma$, where $\Omega$ is a finite domain in $\mathbb{R}^2$, and $\Sigma$ a point set of $N$ points in $\mathbb{R}^2$, which defines the quasilattice. The domain coordinates $(x, y)$ are defined with respect to the coordinate system in Fig. 1, and thus measure the distance relative to the center of Tile(1,1). The center coordinates of Tile(1,1) are the average *x* and *y* coordinates of the 14 vertices. Some points in $\Omega$ with special meaning are the center point, $p_0 = (0,0)$, and the vertices of Tile(1,1), $p_1$ to $p_{14}$. The first vertex has an inner angle of $180°$ [1].

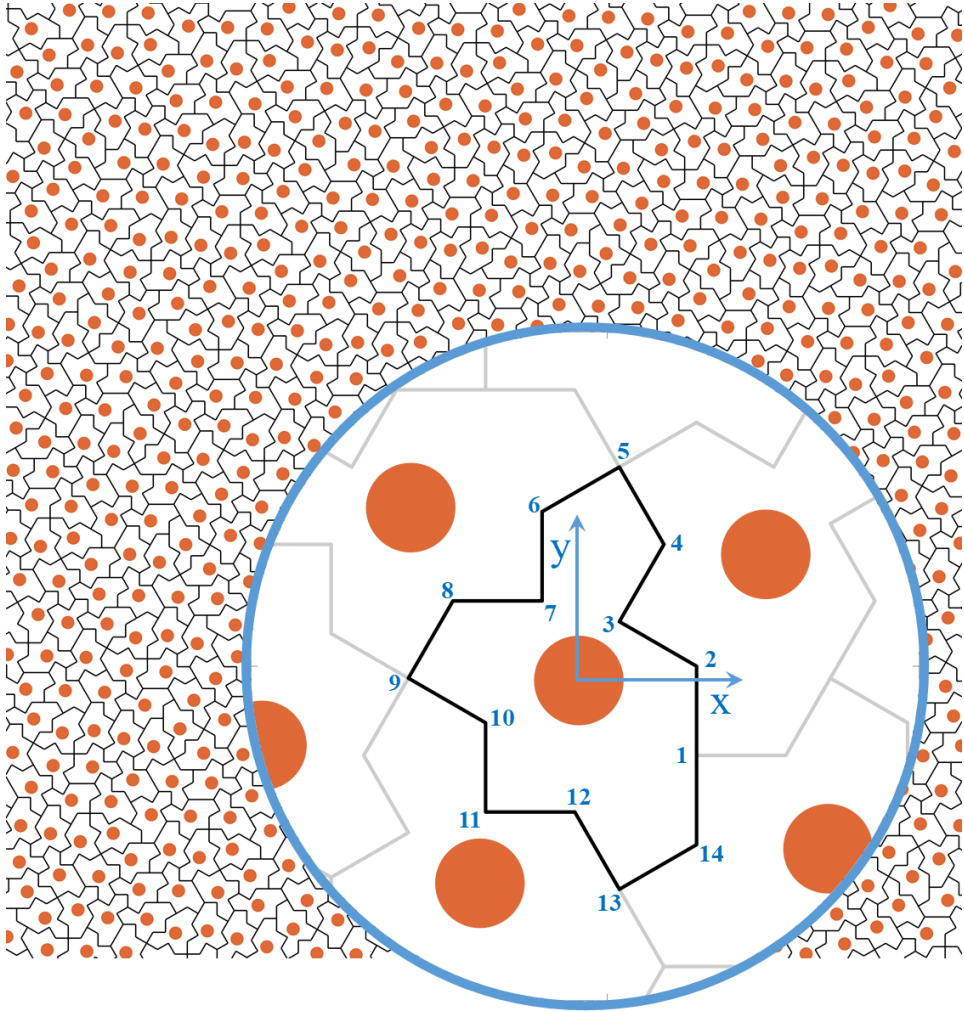

FIG. 1: Tiling with the aperiodic monotile Tile(1,1). In addition to the edges of Tile(1,1), a decoration consisting of disks marking the tile centers is shown ($p_0 = (0,0)$). The disks have a diameter equal to the size of the edges of Tile(1,1), each of which are of unit length. The magnification inset highlights the first tile used in the composition of the Tile(1,1) tiling. Tile(1,1) in this orientation is also used to define the single-tile coordinate system of the lattice generating function. Tile(1,1) vertices are numbered. They are identical to the vertices of all possible Spectre tiles.

Figure 2 shows examples for quasilattices resulting from point decorations located at different coordinates in and around Tile(1,1). Point sets are again shown by small disks rather than points to visually enhance the lattice pattern. The examples have been chosen by visually inspecting the output of the generating function $P(\Omega)$ with $\Omega = [-3, 3] \times [-3, 3]$ for conspicuous patterns exhibiting clear long-range ordering. When their corresponding coordinates were within a distance of about 0.1 to one of the vertices of Fig. 1 (where all edges are of unit length), those were finally selected (points $p_1$ to $p_4$, $p_7$, $p_9$, $p_{10}$). Points $\alpha$, $\beta$, $\gamma$, and $\delta$ are arguments of the generating function that lie outside of Tile(1,1).

Overall, the chosen points give an overview of qualitatively different lattices that can be found in the set $P(\Omega)$. The following observations are notable when the decoration is displayed as small disks: $P(p_0)$ appears as the densest lattice example. $P(p_1)$ consists of clusters. Physically, if the disks denoted the extent of a particle potential, particles would be confined to these clusters. $P(p_2)$ is fully connected, and particles would not be localized anymore. $P(p_3)$ is a combination of a connected part and disconnected parts, and so on. As for arguments outside of Tile(1,1), $P(\alpha)$ and $P(\gamma)$ look relatively sparse, and $P(\delta)$ resembles a noisy hexagonal lattice. In the following, the lattice generating function is further investigated.

A measure of spatial dispersion in a pattern, which will be applied to quasilattices here, is the distance between nearest neighbors (NN). Figure 3A shows the NN distance of points in the lattices $P(\Omega)$ with $\Omega = [-4, 4] \times [-4, 4]$, or $P(x,y)$ for a range of decoration coordinates $x$ and $y$ in $[-4, 4]$ each. Points close to the center of Tile(1,1) (or any Spectre) can never get close to other points (warm colors), but points corresponding to the vertices of Tile(1,1), and many outside points, often do (cold colors). One can see that some of the lattices associated with the vertices of Tile(1,1) have small NN distances.

To quantify this further, we use the 1-NN entropy [5], defined as the average of the logarithm of the distance between nearest neighbor points, up to data-independent terms that are neglected here. This entropy definition avoids the partitioning of the plane for applications like this one where the pattern consists of points. Figure 3B shows that many points with small NN distance also have low 1-NN entropy. The most conspicuous point happens to coincide with point α, whose lattice is relatively sparse. The reason for this sparsity and the low entropy is provided in Fig. 3C, which shows the nine tiles of the first iteration of the tiling of the plane. Seven of the nine decoration points happen to merge into only three common points, thus reducing the density of the lattice and lowering its entropy. One can show (see supplementary code) that the precise value of α is $\frac{[-27\sqrt{3}-31,\ \sqrt{3}-43]}{28} \approx -[2.7773, 1.4739]$.

Interestingly, the sparseness of lattices $P(\gamma)$ and $P(\delta)$ occurs for other reasons, and will be investigated now. Figure 4 shows the $P(\delta) = P(-3, 1.8)$ lattice of Fig. 2 on a larger scale. It resembles a noisy hexagonal lattice that is slightly tilted against the coordinate system of Fig. 1 by an angle $\theta \approx -2.7°$. The statistical six-fold rotational symmetry of the lattice (meaning that each patch occurs in all six orientations with equal frequencies [6]) is also seen in the diffraction [7], which shows six-fold rotational symmetry, too. In fact, it has been shown before that the Fourier transform of the Spectre tiling is non-periodic with chiral six-fold point symmetry about the diffraction origin [8]. Diffraction patterns were computed from the Fourier transform of delta-distributions centered at the decorations (or as the sum of the scattered waves on all the point scatterers), from a circular region to suppress finite domain artefacts [9]. For details, we refer to the supplementary code. As the lattice is tilted when constructed from the rules in Ref. [1], so is the diffraction. The arrow in the inset of Fig. 4 points at one of the six spectral peaks that are responsible for the hexagonal shape of the overall pattern. From this peak, a more accurate estimate of $\theta = -2.7263$ is obtained. As there are many more spectral components in addition to the peaks corresponding to a hexagonal lattice, the overall lattice appears noisy and remains non-periodic. The spatial extension of the hexagonal unit cell corresponding to the marked spectral peak and its symmetric companions is $2\pi/k \approx 7.93$. For comparison, Tile(1,1) has a maximum extension of 4.73. The first iteration of the tiling algorithm for δ is shown in Fig. 3D, revealing that point δ moves the decorations towards the edge of $P(\delta)$. As $P(\delta)$ is the template for larger tilings, the sparse region in its center is copied into the larger tilings, resulting in the cell-like appearance of the pattern in Fig. 4.

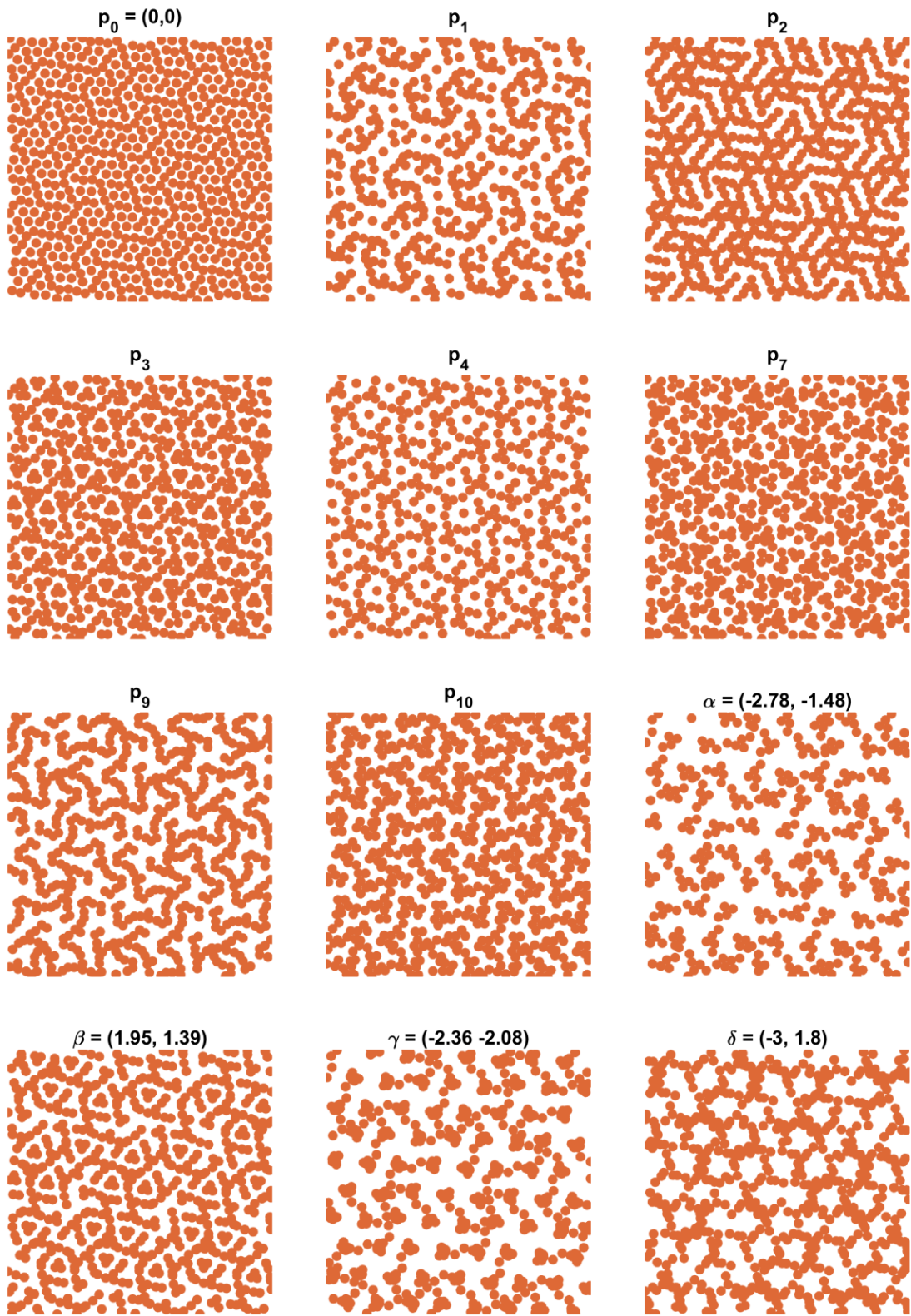

FIG. 2: Example lattices resulting from the same Tile(1,1) tiling but different generating functions $P(x, y)$, with arguments $x, y$ displayed at the top of each pattern. The point sets defining the lattices are shown as small disks. The first eight patterns result from the center point and some of the vertices of Tile(1,1) (Fig. 1). The last four patterns are generated by points that lie outside of Tile(1,1).

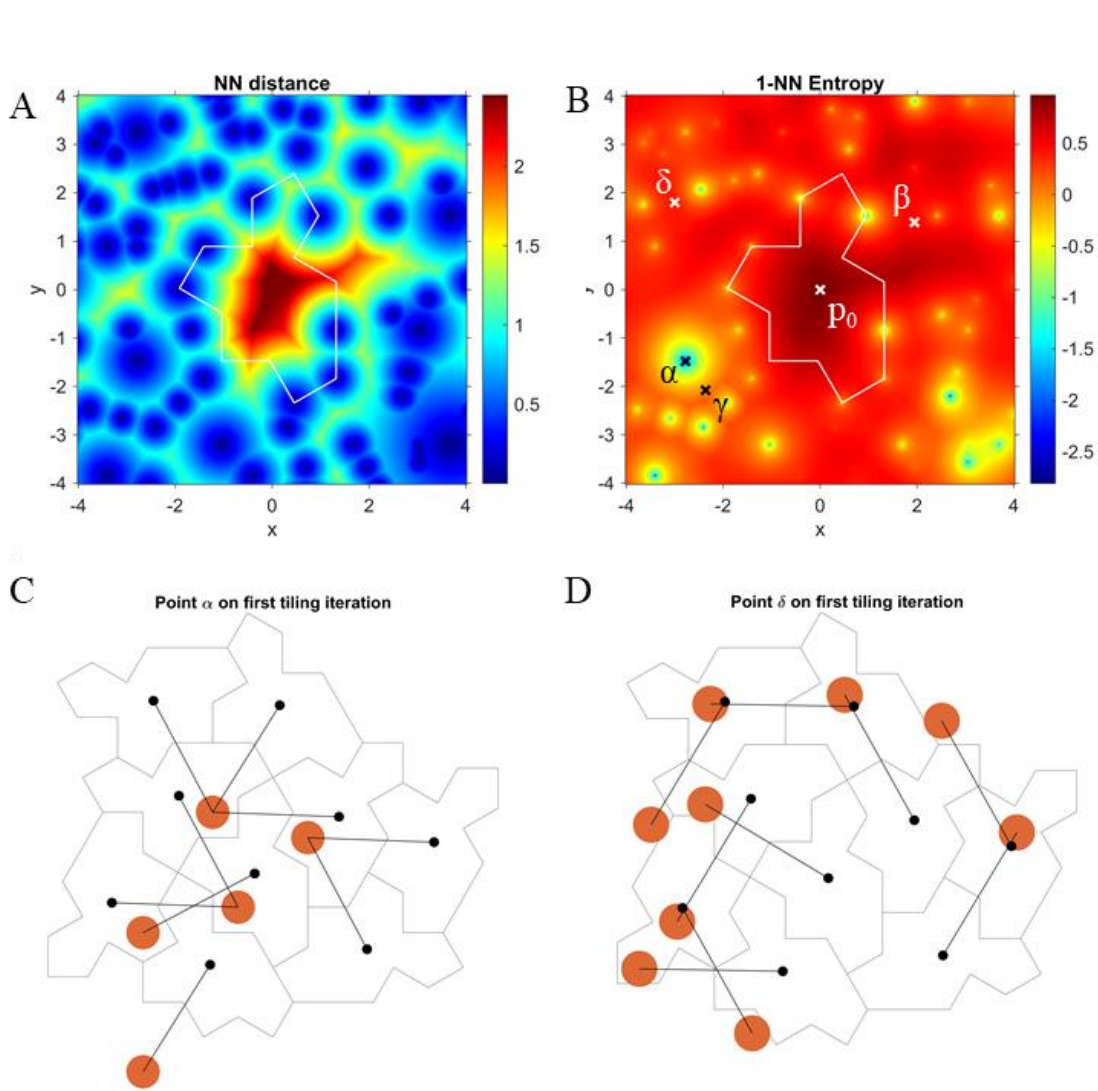

FIG. 3: Nearest neighbor analysis of lattices. A) The minimum nearest-neighbor distance found in lattices $P(\Omega)$ with $\Omega = [-4, 4] \times [-4, 4]$. B) The 1-NN entropy. The largest minimum (blue color) is located at point $\alpha$, whose lattice is shown in Fig. 2. The other special points are annotated, too, but do not correspond to extrema in the entropy. C) For point $\alpha$, black dots denote tile centers, the lines show their relation to point $\alpha$. D) Same for point $\delta$.

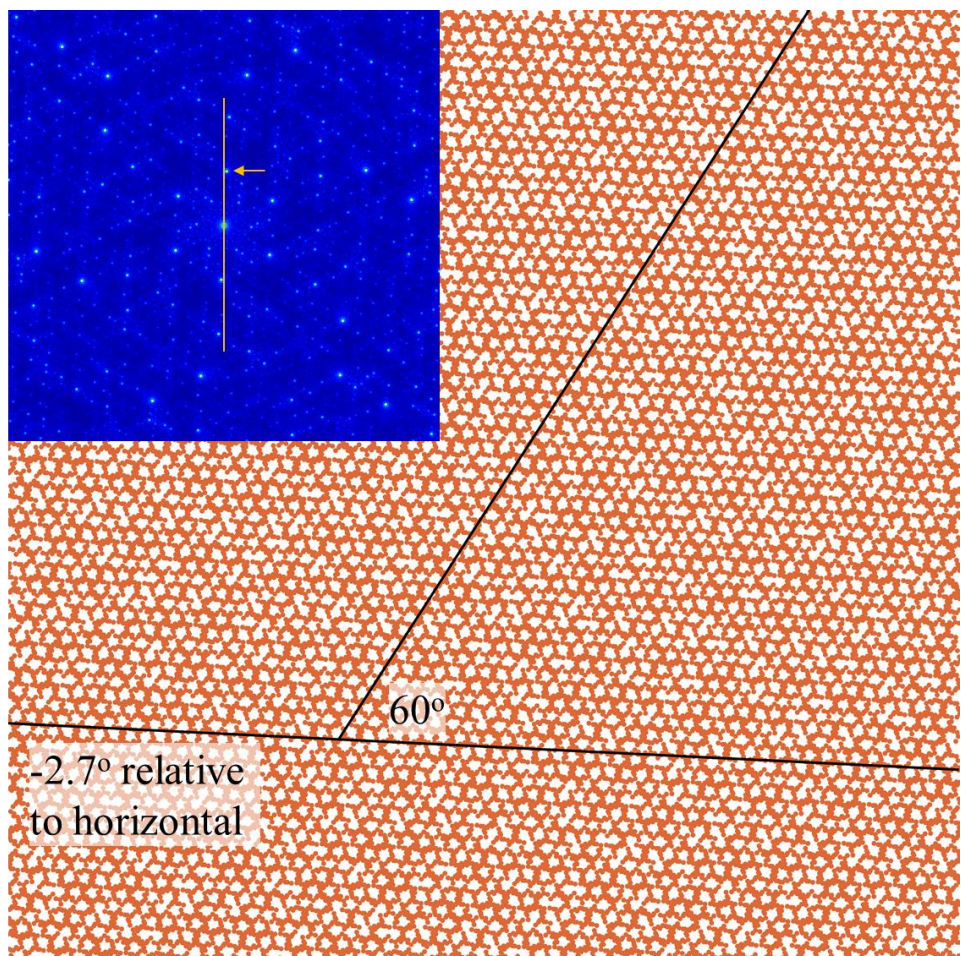

FIG. 4: The $P(\delta)$ quasilattice. On a coarse scale, a hexagonal lattice emerges that is tilted by $\theta \approx -2.7°$ relative to the horizontal axis, as indicated by the bottom black line. The other line at a relative angle of $60°$ and the diffraction pattern (inset) prove the hexagonal statistical symmetry of the quasilattice. The tilt angle $\theta$ transfers into the diffraction pattern; it appears tilted by the same angle relative to the vertical orange line that has been inserted to guide the eye. The arrow points at one of the six spectral peaks responsible for the approximate hexagonal symmetry and the observed average size of the hexagonal cells of about 7.93.

Is $P(\delta)$ the only quasilattice with a strong hexagonal component on this scale? A systematic analysis of the spectral amplitude of the marked peak in the inset of Fig. 4 is shown in Fig. 5A. It seems that $\delta$ lies in a region with high spectral amplitude of this point.

Below $\delta$ there is another region with high spectral amplitude, but the hexagonal structure is not as apparent when looking for example at quasilattice $P(\gamma)$, which lies inside that region. The quasilattices are now rotated by an angle of $-2.7263°$ and their projection onto the vertical axis is investigated. The projection histogram of $P(\gamma)$ (Fig. 5D,E) shows a marked periodicity if compared to the projection of $P(0,0)$ (Fig. 5F,G), which has a more homogeneous histogram. The amplitudes of the main spectral peaks of those projections are shown for the $\Omega = [-4, 4] \times [-4, 4]$ parameter plane in Fig. 5B, and $\gamma$ corresponds to the overall maximum, with a value of 0.48 (relative to a variance of the normalized power spectrum of 1.0). Therefore, lattice $P(\gamma)$ can be explained by a significant projection periodicity along its statistical symmetry axes. In contrast, the lattice $P(p_o)$ has a projection periodicity of only 0.014. The projection periodicity for $P(\delta)$ is 0.31. Finally, this projection periodicity extends to a larger domain $\Omega$ than used before; Fig. 5C shows $P(\Omega)$ for $\Omega = [-25, 25] \times [-25, 25]$, and the projection periodicity itself now has an approximate six-fold symmetry.

The quasilattice $P(\beta)$ appears to be the opposite case to $P(\gamma)$; its spectral amplitude is very low (Fig. 5A), making this lattice more random or non-periodic, in agreement with how it appears visually in Fig. 2.

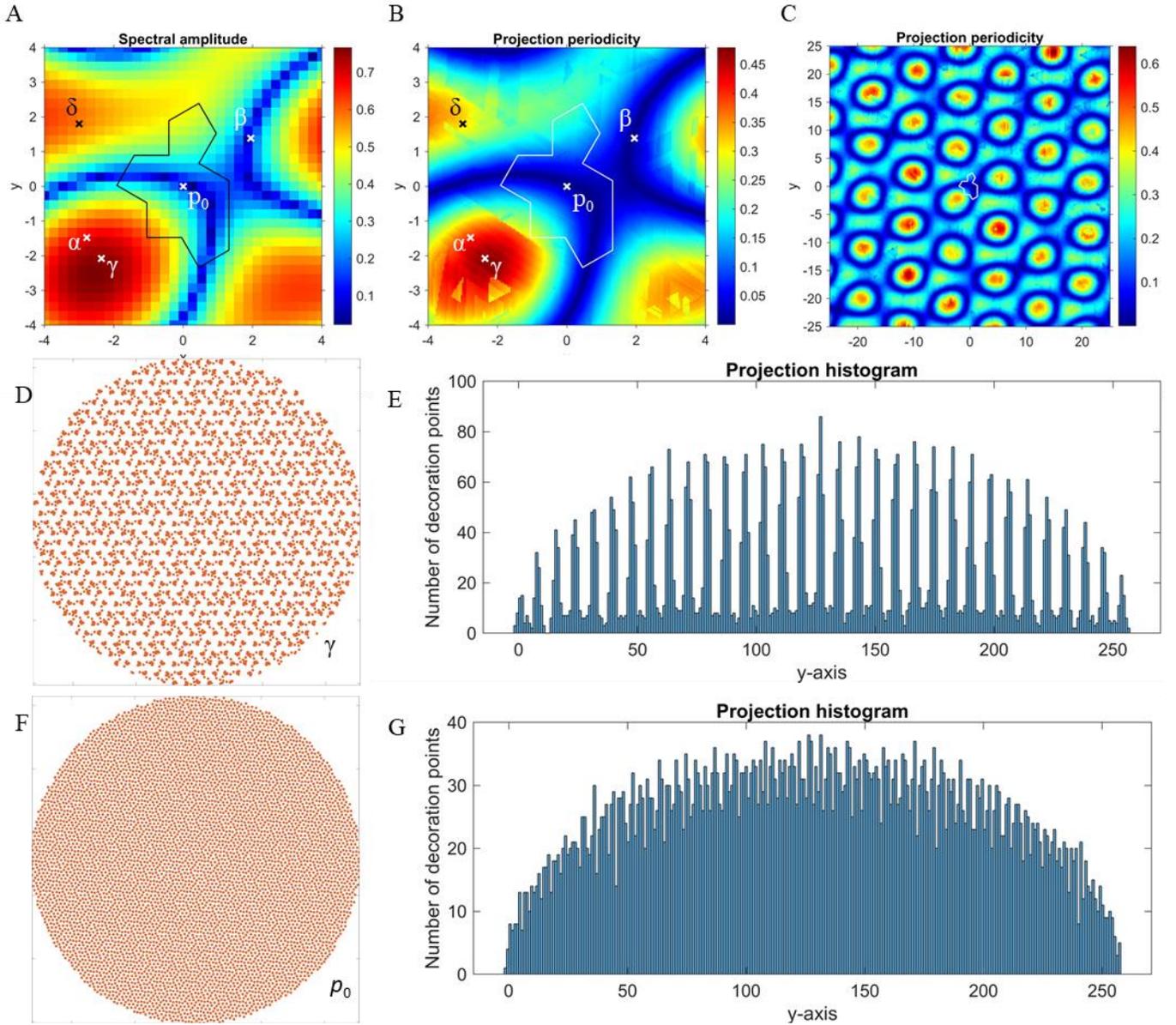

FIG. 5: Spectral analysis. A) The amplitudes of the peak marked with an arrow in the inset of Fig. 4, for the lattice generating function $P([-4, 4] \times [-4, 4])$. B) The projection periodicity of the quasilattices after counterclockwise rotation by 2.7°. C) The same for a larger coordinate range to demonstrate approximate six-fold symmetry of the lattice generating function. D), E) The $P(\gamma) = P(-2.36, -2.08)$ quasilattice and its projection histogram. F), G) For comparison, the $P(p_0)$ quasilattice, which has only a small projection periodicity (also cp. panel B).

Non-periodic tilings by aperiodic monotiles are fundamentally different from periodic tilings or Bravais-lattices. Depending on the decoration of the tile, different quasilattices can be generated, which we have investigated by the lattice generating function. Its argument is the position of the point decoration, and its output is a quasilattice. In periodic tilings, the lattice generating function would be trivial; for all arguments, the lattice would not change and only shift in space. As we have shown, this is not the case for the aperiodic Spectre monotile.

The quasilattices resulting from the lattice generating functions have been regarded as point patterns, and small disks have been used for visualization only. These disks can also have a physical meaning; for example, they could indicate the extent of a potential. Such a system then would be described by a Hamiltonian consisting of non-periodic potentials. If the Spectre could be synthesized or found in nature, this could lead to interesting physical or computational properties, based on previous findings; for example, in a tight-binding model of the Hat monotile, graphene-like but chiral spectral properties have been found [10], and in a quantum dimer model of the Spectre, a deconfined phase has been observed [11]. The synthesis of the Spectre does not seem out of reach considering existing technology for the synthesis of quasicrystals [12-15]. Towards this end, it would be necessary to investigate the band structure of these tilings, and how they relate to the lattices. This could lead to novel methods of band structure tuning. A challenge is that a reciprocal lattice is not well defined, because there are more lattice vectors than dimensions, so they are not all linearly independent. For example, in the Spectre tiling, tiles recur in multiples of 30º rotations. Coupled radiofrequency resonators could be realized relatively easily from resonators made of aperiodic tiles with potential benefits for the design of novel MRI resonators [16, 17]. In this manifestation, coupling between elements depends on their distance, and due to our finding that some quasilattices show a wide range of nearest-neighbor distances, it can be expected that the spectral range can be quite large for different arguments of the lattice generating function.

Recently discovered aperiodic monotiles tile the plane in a non-periodic way by using only one single tile repeatedly. The potential applications [7, 10, 11, 18-23] of these novel parameterizations of the plane inspired us to investigate the quasilattices resulting from different decorations, or markers, on the tilings. By means of a lattice generating function defined by single-point decorations of the Spectre, we investigated the variety of quasilattices that can be generated from one and the same tiling, and provided first attempts at an understanding of this variety. We found that some quasilattices are associated with a low spatial entropy due to their nearest-neighbor properties, and that others show an approximate hexagonal symmetry and approximate periodicity superimposing their general non-periodicity. This latter feature arises after slightly rotating the tiling and finding an approximate periodic structure of its projection onto the symmetry axes of the plane.

*Supplementary information.* The tiling algorithm has been published previously [24]. The analytic derivation of the precise value of α and the diffraction calculation can be found in the annotated MATLAB code for some of the figures of this manuscript. It can be obtained from github.com/henningle/TileOneOne_Quasi.

*AI disclaimer.* We used ChatGPT (GPT-4, OpenAI, 2023, retrieved from https://chat.openai.com) to assist with obtaining ideas about how to perform pattern analysis and with optimizing in-house Matlab code for efficiency. All outputs were manually verified for accuracy and relevance.

---